\documentclass[doublecol]{arXiv} 
\usepackage{graphicx}
\usepackage{epstopdf}
\usepackage{bm}
\usepackage{subfigure}
\usepackage{sidecap}
\usepackage{amsmath}

\usepackage{ltxgrid}
\usepackage{lipsum}

\newcommand{\si}{\sigma}
\newcommand{\al}{\alpha}

\newcommand{\la}{\lambda}

\newcommand{\om}{i\omega_n}
\newcommand{\Om}{\Omega}
\newcommand{\ep}{\epsilon}
\newcommand{\varep}{\varepsilon}
\newcommand{\bxi}{\bar{\xi}}

\newcommand{\bsig}{{\bm\sigma}}

\newcommand{\blam}{{\bm\lambda}}

\newcommand{\hblam}{\hat{{\bm\lambda}}}

\newcommand{\be}{\begin{equation}}
\newcommand{\ee}{\end{equation}}

\newcommand{\bea}{\begin{eqnarray}}
\newcommand{\eea}{\end{eqnarray}}
\newcommand{\bd}{\begin{displaymath}}
\newcommand{\ed}{\end{displaymath}}
\newcommand{\ba}{\begin{array}}
\newcommand{\ea}{\end{array}}
\newcommand{\bi}{\begin{itemize}}
\newcommand{\ei}{\end{itemize}}
\newcommand{\bc}{\begin{center}}
\newcommand{\ec}{\end{center}}
\newcommand{\bfl}{\begin{flushleft}}
\newcommand{\efl}{\end{flushleft}}
\newcommand{\bfr}{\begin{flushright}}
\newcommand{\efr}{\end{flushright}}

\newcommand{\hG}{\hat{G}}

\newcommand{\tL}{\tilde{\Lambda}}

\newcommand{\ts}{\tilde{s}}
\newcommand{\te}{\tilde{\varep}}
\newcommand{\tE}{\tilde{E}}

\newcommand{\rar}{\rightarrow}

\newcommand{\SP}{S\lowercase{r}P\lowercase{t}A\lowercase{s}}

\newcommand{\bl}{\begin{aligned}}
\newcommand{\el}{\end{aligned}}

\def\dg{^{\dagger}}

\def\bk{{\bf k}} \def\bq{{\bf q}}

 \def\bd{{\bf d}}

\def\da{\downarrow} \def\ua{\uparrow}
 \def\rar{\rightarrow}
\def\dg{\dagger}

\def\={\!\!\!&=&\!\!\!}
\def\+{\!\!\!&&\!\!\!+~}
\def\-{\!\!\!&&\!\!\!-~}

\title{Gap function of hexagonal pnictide superconductor SrPtAs from quasiparticle interference spectrum}

\author{Alireza Akbari\inst{1} 
\and Peter Thalmeier\inst{2}
}
\shortauthor{A. Akbari \etal}

\institute{                    
  \inst{1}
  Max Planck Institute for Solid State Research, D-70569 Stuttgart, Germany \\
     \inst{2}
   Max-Planck Institute for the  Chemical Physics of Solids, D-01187 Dresden, Germany \\
  }
\pacs{74.20.Rp}{Pairing symmetries}
\pacs{74.55.+v}{Tunneling phenomena: single particle tunneling and STM}
\pacs{74.70.Xa}{Pnictides and chalcogenides}

\abstract{
The  pnictide superconductor \SP~ has a hexagonal layered structure containing inversion symmetry. It is formed by stacking two inequivalent PtAs layers separated by Sr layers. The former have no local (in-plane) inversion symmetry and therefore a (layer-) staggered Rashba spin orbit coupling appears which splits the three Kramers degenerate bands into six quasi-2D bands.
 The symmetry of the superconducting state  of \SP~ is unknown. Three candidates, spin-singlet $A_{1g}$ and $E_g$ as well as triplet $A_{2u}$ states have been proposed.  We predict the quasiparticle interference (QPI) spectrum for these gap functions in t-matrix Born approximation. We show that distinct differences in the pattern of characteristic QPI wave vectors appear. These results may be important to determine the gap symmetry of \SP~ by STM-QPI method.
 }

\begin{document}

\maketitle

\twocolumngrid

Transition metal pnictide superconductors (SC) in particular the Fe-based systems are all of the tetragonal (orthorhombic) structure. 
The layered Pt-pnictide \SP \cite{nishikubo:11}  is the first superconductor  ($T_c= 2.4$ K) in that class with hexagonal structure composed of honeycomb Pt-As layers spaced by Sr layers.
It may be viewed as a MgB$_2$ type structure with Mg sites ocuppied by Sr and B sites in an ordered fashion such that Pt-As alternate in the 2D honeycomb layers as well along the hexagonal c-axis.
The resulting structure has an overall 3D inversion center whereas the individual layers lack 2D inversion symmetry which is not contained in the $C_{3v}$ layer point group.

Because the electronic states at the Fermi level are mostly of Pt(5d) type with strong spin orbit coupling this leads to a peculiar 
electronic band structure \cite{youn:12a}. Firstly the two inequivalent Pt-As layers have only small interlayer hopping which results in a 
quasi-2D band structure consisting of three hole bands and associated Fermi surface (FS) columns. Secondly an effective 2D Rashba spin orbit coupling term leads to a large splitting of the three bands which depends on $k_z$ in such a way that overall 3D inversion symmetry is restored.

This has consequences for the possible superconducting pair states. Due to essentially decoupled layers it is reasonable to assume only intra-layer pairing. Then one can expect features as in the non-centrosymmetric superconductors consisting of a mixture of spin-singlet and triplet pairing of the in-plane order parameter due to lack of local 2D inversion symmetry. For the overall 3D superconducting state even or odd parity classification is restored due to the two inequivalent Sr-Pt layers.  The  momentum dependence of these unconventional pair states was investigated theoretically by Goryo et al \cite{goryo:12} and it was found that even A$_{1g}$, E$_g$ and odd A$_{2u}$ states are viable candidates. However sofar there is only few experimental evidence to discriminate between them \cite{matano:14}.\\

One of the most powerful recent methods to determine the symmetry of the gap function is STM quasiparticle interference (QPI) technique \cite{capriotti:03}. The Fourier transform of the differential conductance scans as function of bias voltage give a fingerprint of the Fermi surface in the normal state and in addition of the {\bf k}- dependence of the gap function in the SC state. It has by now been successfully applied to a variety of cuprate\cite{mcelroy:2003,hanaguri:2009,Hoffman:2002,Wang:2003,pereg:08, maltseva:2009,balatsky:06}, Fe-pnictides \cite{Hanaguri:2010,allan:12,Chuang:2010,zhang:09,Akbari:2010,Knolle:2010,huang:11}, 
and heavy fermion unconventional superconductors \cite{akbari:11,allan:13,zhou:13}.
There are no STM results yet for the hexagonal pnictide SC \SP.\\
Therefore in this work we propose the application of QPI to investigate the \SP~ SC gap function. We will compare the predicted QPI spectra for the three main gap candidates discussed sofar to provide criteria for discriminating among them  in future STM experiments.

%
\begin{figure}[t]
\includegraphics[width=0.98\linewidth]{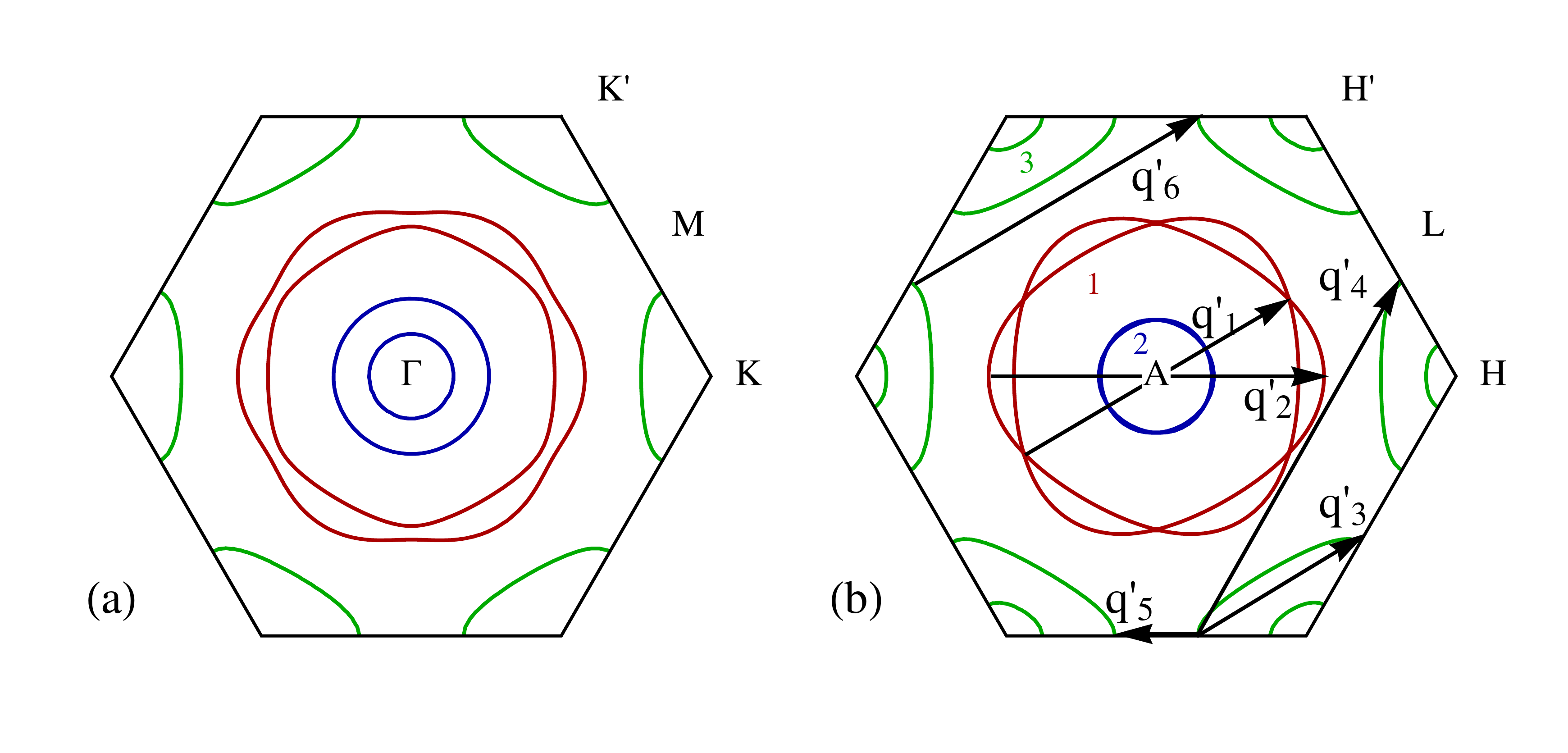}
\caption{(Color online)
Fermi surface cuts through symmetry planes $k_z=0$ (a) and $k_z=\pi/c$ (b). b=1-3 denotes three pairs of conduction bands  parametrized according to Eq.~(\ref{eq:kfact}) with parameter set ($t_1^b,t_c^b,t_{c2}^b,\mu_b,\al_b$). We use  \cite{youn:12a} for $b=1$: (1.25, 0.1, 0.05, 0.5, 0.4), for $b=2$: (1.0, 0.1, 0.05, 2.5, 0.28) and for $b=3$: (-0.48, 0.075, -0.03, 0.6, 0.046). $\bq'_i$ are characteristic QPI scattering vectors.}
\vspace{-0.5cm}
\label{fig:Fig1}
\end{figure}
%

The one-body Hamiltonian for three  \SP~ hole bands of mixed As(4p)-Pt(5d) character close to the Fermi energy derived in Ref.~\cite{youn:12a} and used in Refs.~\cite{goryo:12,youn:12b} is given by
\be
H_0=\sum_{\bk ,ll',s,b}\ep_{\bk ll'}^bc_{\bk ls}^{b\dagger}c_{\bk l's}^b +
\sum_{\bk ,l,ss',b}\al_b\blam_{\bk l}\cdot\bsig_{ss'}    c_{\bk ls}^{b\dagger}c_{\bk ls'}^b ,
\label{eq:ham0}
\ee
where $c_{\bk ls}^{\dagger b}$ creates conduction electrons with $b=1,2,3$ denoting the (hole) band, $l,l'=1,2$ the inequivalent PtAs layers and $s=\pm\frac{1}{2}$ the (real) spin.
Furthermore $\ep_{\bk ll'}^b=\varep_{\bk ll'}^b-\mu\delta_{ll'}$  ($\mu$= chemical potential) is the Fourier transformed ($l,l'$=layer) hopping matrix and $\al_b\blam_{\bk ll'}=\al_b\la^z_\bk(-1)^{(l+1)}\delta_{ll'}\hat{{\bf z}}$ the Rashba-type spin orbit coupling matrix for the PtAs layers which lack inversion symmetry. 
They are given by
\be
\hat{\ep}_\bk^b=
\left[
 \begin{array}{cc}
 \varep_\bk^b-\mu_b& \varep_{c\bk} \\
 \varep_{c\bk}^*& \varep_\bk^b-\mu_b
\end{array}
\right],\;\;\;
\hat{\la}^z_\bk=
\left[
 \begin{array}{cc}
 \la^z_\bk& 0 \\
 0& -\la^z_\bk
\end{array}
\right].
\label{eq:bandmat}
\ee
The intra-layer hopping is described by  $\varep^b_\bk$ and the inter-layer hybridization by $\varep^b_{c\bk}$.
Here $\hat{\la}^z_\bk$ has opposite signs for $l=1,2$ to restore the global inversion symmetry. Its strength is given by the orbital (band) dependent  Rashba coupling $\alpha_b$.
Explicitly \cite{youn:12a,goryo:12},
\begin{equation}
 \begin{aligned}
&\varep_\bk^b=t_1^b
\Big
[\cos k_ya +2\cos\frac{\sqrt{3}k_x a}{2}\cos\frac{k_y a}{2}
\Big]
+t_{c2}^b\cos k_zc
\\&
|\varep_{c\bk}^b|^2=t_c^{b2}\cos^2\frac{k_zc}{2}
\Big[
3+2\cos k_ya+4\cos\frac{\sqrt{3}k_x a}{2}\cos\frac{k_y a}{2}
\Big]
\\&
\la^z_\bk=\sin k_ya - 2\cos\frac{\sqrt{3}k_x a}{2}\sin\frac{k_y a}{2}.
\label{eq:kfact}
 \end{aligned}
\end{equation}
The hopping and Rashba parameters for realistic Fermi surface hole sheets \cite{youn:12b}  are given in  Fig.~\ref{fig:Fig1}. From $H_0$ the normal state quasiparticle bands are
\be
\Omega_{\bk\pm}^{b}=(\varep^b_\bk-\mu_b)\pm\sqrt{|\varep_{c\bk}^b|^2+\al_b^2\la^{z2}_\bk}.
\label{eq:qpnormal}
\ee
The Fermi surface cuts of the six bands (b=1-3,$\pm$) which are twofold Kramers (pseudo-spin) degenerate are shown in Fig.~\ref{fig:Fig1} for the normal state. The difference between $k_z=0, \pi/c$ is due to the effect of interlayer hopping $\varep^b_ {c\bk}$.\\

\begin{figure}[t]
\vspace{-0.5cm}
\includegraphics[width=1. \linewidth]{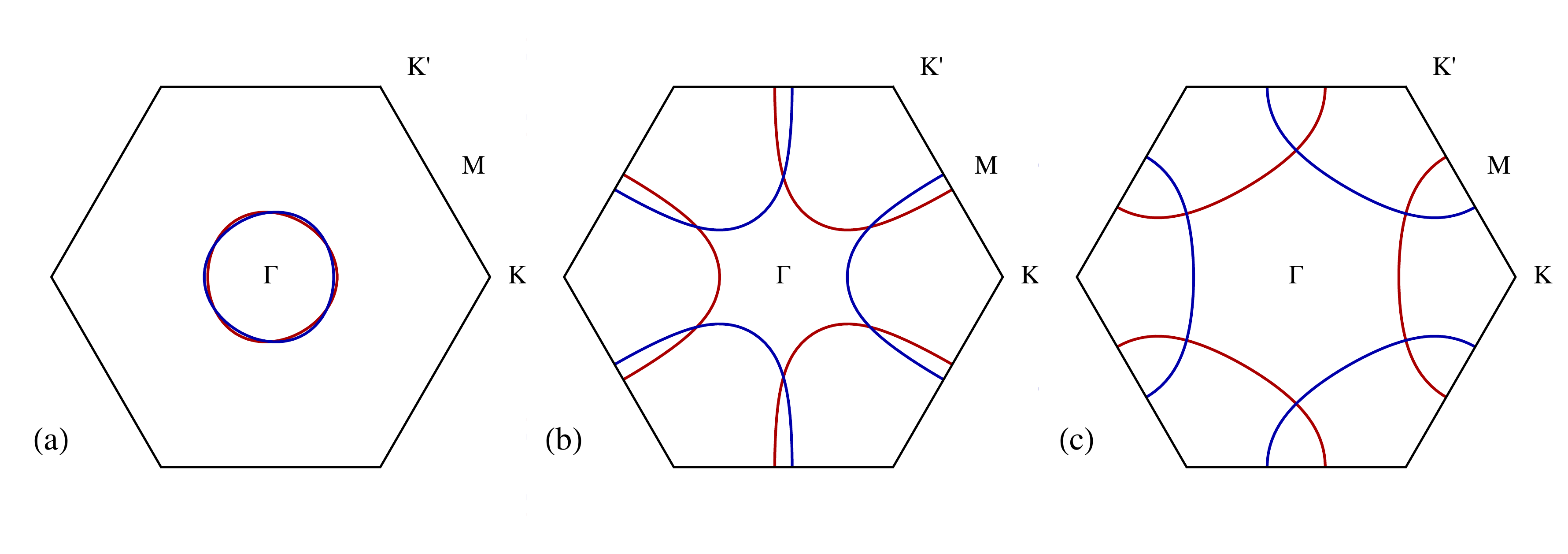}
\includegraphics[width=1. \linewidth]{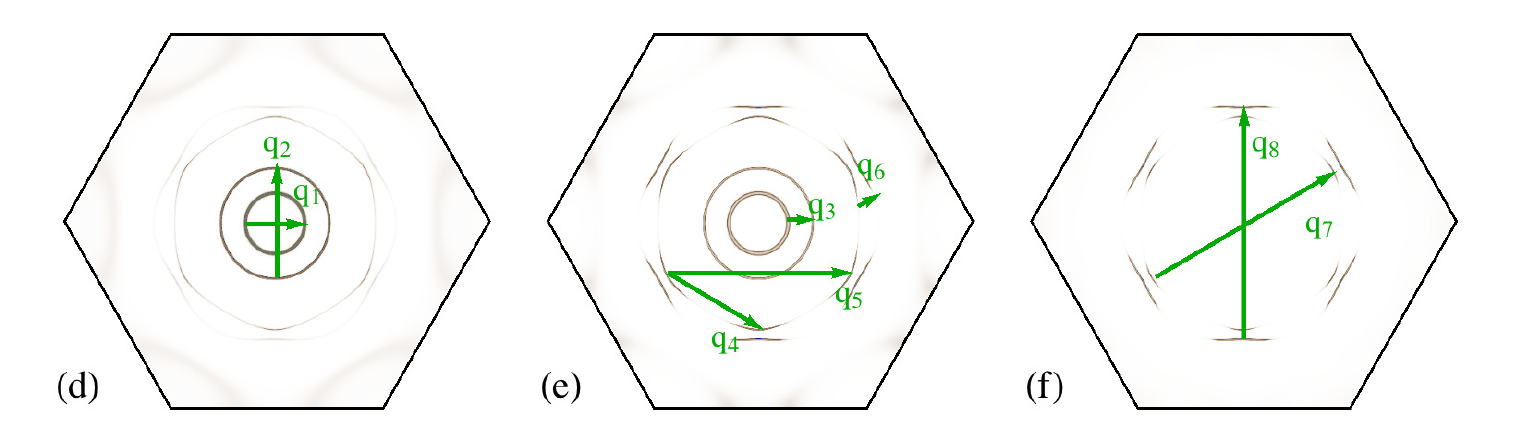}
\caption{(Color online)
Node structure of the (a) $A_{1g}$, (b) $A_{2u}$ and (c) $E_g(1,1)$ gap functions $\Delta^b_{\bk\pm}$ in Eq.~(\ref{eq:scgap}) ($+$ and $-$ correspond to red and blue, respectively).
Equal SC quasiparticle energy (Eq.~(\ref{eq:qpscD2})) surfaces for $\omega/\Delta_0^b= 0.5$  for each of the three ($b=1-3$) gap functions (d-f). The characteristic wave vectors $\bq_i$ connecting maximum curvature points are indicated. 
For the nondegenerate representations $\Delta^b_{\bk\pm}$  are parametrized by $\Delta_b^0$ and two additional parameters, namely for $A_{1g}$: $(s_b,t'_b)=(-0.51,0.12)$ and for $A_{2u}$: $(\ts_b,s_b)=(0.15,-0.18)$. For the degenerate $E_g(1,1)$ we use $t'_b=1$. The gap amplitude  was chosen  as $\Delta_b^0 = 0.05$.
}
\label{fig:Fig2}
\end{figure}

Possible superconducting gap functions were proposed in Refs.~ \cite{fischer:11,goryo:12,Fischer:14,youn:12b}. The most likely candidates are the even singlet $A_{1g}$ and $E_g$ and the odd triplet $A_{1u}$ representations. Their explicit \bk - dependence on the six bands is given by
\bea
 \begin{aligned}
A_{1g}:\;\;\Delta^{b}_{\bk\pm}&=&\Delta_0^b(1+s_be_\bk\pm t'_bh_\bk)
\\
A_{2u}:\;\;\Delta^{b}_{\bk\pm}&=&\Delta_0^b(\tilde{s}_b+s_be_\bk\pm h_\bk)  
\\
E_{g}\;:\;\;\Delta^{b}_{\bk\pm}&=&\sum_{s=\pm}\Delta^b_{0s}(e^s_{\bk s}\pm t'_{b}h^s_{\bk})
\label{eq:scgap}
 \end{aligned}
\eea
where $s=\pm$ denotes the time reversed chiral states of $E_{g}$ with $e^\pm_\bk=e_{\bk 1}\pm ie_{\bk 2}$ and  $h^\pm_\bk=h_{\bk 1}\pm ih_{\bk 2}$ (1,2 correspond to real and imaginary parts). Here $\Delta_0^b$  and  $\Delta_{0s}^b (s=\pm)$  are gap amplitudes and $(s_b,\ts_b) ,t'_{b}$ are admixture amplitudes of singlet and triplet parts. They will be assumed as band (b) independent in agreement with microscopic considerations \cite{goryo:12}.
We restrict the twofold degenerate $E_{g}$ manifold to $E_{g}(1,1)$ with $\Delta^b_{0s}=\Delta^b_0$. For simplicity we do not consider the chiral state $E_{g}(1,i)$ \cite{Fischer:14} which breaks time reversal symmetry \cite{biswas:13}.
Then $\Delta^b_{\bk\pm}$ may be chosen real. This leads to
\be
E_{g}(1,1)\;:\;\;\Delta^{b}_{\bk\pm}=2\Delta^b_{0}(e_{\bk 1}\pm t'_{b}h_{\bk 1}).
\label{eq:scgapE11}
\ee
The (real) layer gap matrices in spin space ($\uparrow,\downarrow$) are then given by
\be
\hat{\Delta}^b_{\bk 1}=
\left[
 \begin{array}{cc}
 0& p\Delta^b_{\bk -} \\
-p\Delta^b_{\bk +} & 0
\end{array}
\right],
\;
\hat{\Delta}^b_{\bk 2}=
\left[
 \begin{array}{cc}
 0& \Delta^b_{\bk +} \\
-\Delta^b_{\bk -} & 0
\end{array}
\right],
\label{eq:scmat}
\ee
with p denoting the gap parity $p=1$ for $A_{1g}, E_g$ and $p=-1$ for $A_{2 u}$. The form factors in Eqs.~(\ref{eq:scgap},\ref{eq:scgapE11}) are defined by
\bea
 \begin{aligned}
e_\bk&=\cos k_ya+2\cos\frac{\sqrt{3}k_xa}{2} \cos\frac{k_ya}{2},
\\
h_\bk&=\sin k_ya-2\cos\frac{\sqrt{3}k_xa}{2}\sin\frac{k_ya}{2},
\label{eq:aform}
 \end{aligned}
\eea
for the nondegenerate ($A_{1g}$ and $A_{2u}$) case and for twofold degenerate $E_{g}$ gap function we have
\be
 \begin{aligned}
e_{\bk 1}&=-\frac{1}{2} e_{\bk },\;\;
h_{\bk 1}=-\frac{1}{2} h_{\bk },
\\
%
e_{\bk 2}&=\frac{\sqrt{3}}{2}[\cos k_ya -\cos\frac{(\sqrt{3}k_x-k_y)a}{2}],
\\
h_{\bk 2}&=\frac{\sqrt{3}}{2}[\sin k_ya -\sin\frac{(\sqrt{3}k_x-k_y)a}{2}].
\label{eq:eform}
 \end{aligned}
\ee
Due to even (e) and odd (h) form factors the gap elements in  Eqs.~(\ref{eq:scgap},\ref{eq:scmat}) fulfill the relation $\Delta_{-\bk\pm}=\Delta_{\bk\mp}$.
Therefore under inversion $ I(\bk, 1)=(-\bk,2)$,  $I(\bk,2)=(-\bk, 1)$ the real  layer gap matrices in Eq.~(\ref{eq:scmat}) exhibit the proper even ($p=1$) or odd ($p=-1$) symmetry $\hat{\Delta}^b_{\bk 1}=p\hat{\Delta}^b_{-\bk 2}$ and 
$\hat{\Delta}^b_{\bk 2}=p\hat{\Delta}^b_{-\bk 1}$.
 The nodal structure of these gap functions is shown in Fig.~(\ref{fig:Fig2}a-c).\\

%
%

Adding the pairing term to $H_0$ which includes only intra-band and -layer terms this leads to a BCS model 
\be
H_{SC}=H_0+
\frac{1}{2}\sum_{\bk l ss' b}(\Delta_{\bk l }^{b ss'}c_{-\bk l s}^{b\dg} c_{\bk l s'}^{b\dg} +H.c.).
\label{eq:HBCS}
\ee
It is associated with the (inverse) Green's function matrix $\hG^{-1}=\hG^{-1}_A\otimes\hG^{-1}_B$ with $\hG^{-1}_{A,B}=(\om-H^{A,B}_{SC})$ . Suppressing the band index (b) for the moment and expressing $H^{A,B}_{SC}$ in the spinor basis
$(c^\dg_{\bk 1\ua},c_{-\bk 1\da},c^\dg_{\bk 2\ua},c_{-\bk 2\da})$  for A and $(c^\dg_{\bk 1\da},c_{-\bk 1\ua},c^\dg_{\bk 2\da},c_{-\bk 2\ua})$ for B we have:\\
\newline
\onecolumngrid
\noindent\rule{0.5\linewidth}{0.4pt}
\be
 \begin{aligned}
\hG^{-1}_A({\bk,\om})=
\left[
 \begin{array}{cccc}
\om-\te_\bk-\alpha\la^z_\bk & -p\Delta_{\bk -}& -\varep_{c\bk}&0 \\
-p\Delta^*_{\bk -}& \om+\te_\bk+\alpha\la^z_\bk &0&\varep_{c\bk}\\
 -\varep^*_{c\bk}&0&\om-\te_\bk+\alpha\la^z_\bk& -\Delta_{\bk +}\\
 0 &  \varep^*_{c\bk}& -\Delta^*_{\bk +}&\om+\te_\bk-\alpha\la^z_\bk
\end{array}
\right],
\label{eq:Greensinv}
 \end{aligned}
\ee
\hfill\noindent\rule{0.5\linewidth}{0.4pt}
\twocolumngrid
~\newline
where we define $\te^b_\bk=\varep^b_\bk-\mu_b$ for each band.
Then $\hG^{-1}_B({\bk,\om})$ may be obtained by substituting $\la^z_\bk\rar-\la^z_\bk$ and $\Delta_{\bk\pm}\rar -\Delta_{\bk\mp}$ in the above equation. Note that the model gap functions  $A_{1g}$ and $A_{2u}$ of Eq.~(\ref{eq:scgap}) and $E_g (1,1)$ of  Eq.~(\ref{eq:scgapE11}) are chosen real, i.e., $\Delta_{\bk\pm}^*=\Delta_{\bk\pm}$.\\

After inversion $\hG^(\bk,\om)$ may be used to calculate the QPI spectrum $\tL_0(\bq,\om)$ which is proportional to the spatial Fourier transform of the STM differential conductance \cite{capriotti:03}. We assume that only a \bq - independent non-magnetic impurity scattering $U_c$ is present. For weak scattering with $U_cN_b(\mu)\ll 1$ ($N_b=$ DOS of band b) we may restrict to Born approximation. Even when this is not valid full t-matrix theory gives very similar results for the \bq -space structure of the QPI function \cite {akbari:13a}.
Within  Born approximation \cite{akbari:13,akbari:13b} it is given by $\tL_0(\bq,\om)=U_c\Lambda_0(\bq,\om)$ with (summation over  b is implied)
\be
\Lambda_0(\bq,\om)=\frac{1}{2N}\sum_\bk tr_{\si\tau\kappa}
\Big[\frac{1+\tau_3}{2}\hG_\bk\tau_3\sigma_0\kappa_0\hG_{\bk-\bq}
\Big]. 
\label{eq:QPItrace}
\ee
The trace is performed with respect to Nambu spin  ($\tau$), real spin ($\sigma$) and layer index ($\kappa$) where $\tau_3$ is a Pauli matrix and $\sigma_0,\kappa_0$ are unit matrices.\\

First we discuss the purely 2D model for \SP~ neglecting the dispersion along $k_z$ setting $\varep^b_{c\bk}\equiv 0$. Then the Fermi surface cut for each $k_z$ is equivalent to that of Fig.~\ref{fig:Fig1}b ($k_z=\pi/c$) where $\varep^b_{c\bk}\equiv 0$ vanishes even for the 3D case with finite inter-layer hybridization. In the 2D model the Green's function can be obtained easily by inverting Eq.~(\ref{eq:Greensinv}) due to  $\varep^b_{c\bk}= 0$. 
To perform the traces in Eq.~(\ref{eq:QPItrace}) it is convenient to transform $\hG(\bk,\om)$ to reordered spinor basis 
$(c^\dg_{\bk 1\ua},c^\dg_{\bk 2\ua},c^\dg_{\bk 1\da},c^\dg_{\bk 2\da})$  ($\tau_3=+1$) and 
$(c_{-\bk 1\da},c_{-\bk 2\da},c_{-\bk 1\ua},c_{-\bk 2\ua})$ ($\tau_3=-1$). 
Then the QPI spectrum per spin and layer is obtained from Eq.~(\ref{eq:QPItrace}) explicitly as
%
%
\be
 \begin{aligned}
&\Lambda_0(\bq,\om)=
\\&
\hspace{1cm}
\frac{1}{2N}\sum_{\bk b\xi}\frac
{(\om+\te^b_{\bk\xi})(\om+\te^b_{\bk-\bq\xi})-\Delta^b_{\bk\xi}\Delta^{b*}_{\bk-\bq\xi}}
{[(\om)^2-\tE^2_{\bk\xi}][(\om)^2-\tE^2_{\bk-\bq\xi}]},
\;\;\;
\label{eq:QPI2D}
 \end{aligned}
\ee
%
%
where the band-index $b$ has been reintroduced. Furthermore the branches of superconducting Rashba-split quasiparticle bands are given by ($\xi=\pm$) 
\be
\tE^{b2}_{\bk\xi}=\te^{b2}_{\bk\xi }+|\Delta_{\bk\xi}|^2  ; \;\;\;
\te^{b}_{\bk\xi}=\te_\bk-\xi\alpha_b\lambda_\bk^z.
\label{eq:qpscD2}
\ee
Equation (\ref{eq:QPI2D}) may be used for the calculation of the 2D QPI spectrum provided the model for $\Delta^b_{\bk\pm}$ 
is specified. Here we refer to results obtained previously \cite{akbari:13,akbari:13a} on QPI in truly non-centrosymmetric superconductors with global inversion symmetry breaking. It was found there that generally in the expression for $\Lambda_0(\bq,\om)$ additional Rashba coherence factors of the type $\frac{1}{2}[1+\xi\xi'\hblam_\bk\hblam_{\bk'}]$  with unit vector $\hblam_\bk=\blam_\bk/|\blam_\bk|$ and $\xi,\xi'=\pm$ are present. However in our present case $\blam_\bk=\la_{\bk}^z\hat{{\bf z}}$ has only one component and the coherence factors are just one or zero  and by a suitable definition of the Rashba split bands as in Eq.(\ref{eq:qpscD2})  they do not appear explicitly in Eq.(\ref{eq:QPI2D}).   For the same reason the latter also describes the QPI spectrum for magnetic scattering in Born approximation.\\

%
\begin{figure}[t]
\vspace{0.0cm}
\includegraphics[width=1. \linewidth]{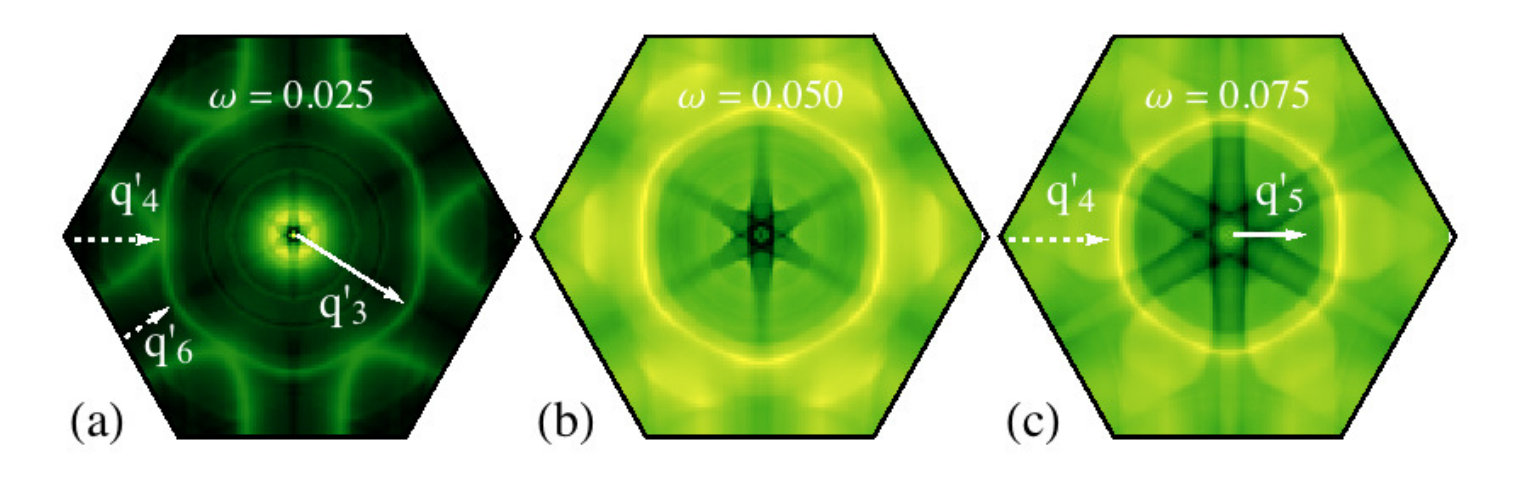}
\caption{ (Color online)
QPI spectrum (average) in the normal state for three different frequencies $\omega$ using the 2D FS model with $t^b_c=0$. Characteristic wave vectors $ \bq'_i$ of the normal state FS in Fig.~\ref{fig:Fig1}b are indicated.}
\label{fig:Fig3}
\end{figure}
%

For the numerical calculation we use the general 3D QPI by including  the interlayer hopping and its resulting $k_z$ dispersion. When $\varep_{c\bk}$ is nonzero  the quasiparticle excitation spectrum is obtained by the  zeroes of the determinant $D_b(\bk,\om)=0$ which is given by
\be
 \begin{aligned}
D_b(\bk,\om)&=
\Big[
(\om)^2-\tE^{b2}_{\bk -}
\Big]
\Big[
(\om)^2-\tE^{b2}_{\bk+}
\Big]
+|\varep^b_{c\bk}|^4
\\
&-2|\varep^b_{c\bk}|^2
\Big[
(\om)^2+(\te^b_{\bk+}\te^b_{\bk-}-p\Delta^b_{\bk +}\Delta^b_{\bk _-})
\Big] .
\label{eq:det}
 \end{aligned}
\ee
The 3D quasiparticle energies $\Om^{b}_{\bk \xi}$, including the effect of interlayer hopping $\varep^b_{c\bk}$ with dispersion along $k_z$, are obtained as
\be
 \begin{aligned}
\label{eq:qpscD3}
\Om^{b2}_{\bk \pm}=&\frac{1}{2}(\tE^{b2}_{\bk+}+\tE^{b2}_{\bk-})+|\varep^b_{c\bk}|^2\pm
\Big[
\frac{1}{4}(\tE^{b2}_{\bk+}-\tE^{b2}_{\bk-})^2
\\
&+|\varep^{b}_{c\bk}|^2[(\te^b_{\bk+}+\te^b_{\bk-})^2+(\Delta^b_{\bk+}-p\Delta^b_{\bk_-})^2]
\Big]^\frac{1}{2}.
\;\;\;
\;\;\;\;\;\;
 \end{aligned}
\ee
Here $\te^b_{\bk+}+\te^b_{\bk-}=2\te^b_{\bk}$. For $\Delta^b_{\bk\pm}=0$ we recover the quasiparticle bands $\Omega^{b}_{\bk \xi}$ of the normal state in Eq.(\ref{eq:qpnormal}).
Obviously for $\varep_{c\bk}=0$ the  $\Om^{b}_{\bk \xi}$ reduce to the  $\tE^{b}_{\bk \xi}$ of Eq.~(\ref{eq:qpscD2}).
Then, after the inversion of Eq.~(\ref{eq:Greensinv}) and performing the trace in Eq.~(\ref{eq:QPItrace}) we obtain the general 3D QPI function as
\newline
\onecolumngrid
\noindent\rule{0.5\linewidth}{0.4pt}
\be
\Lambda_0(\bq,\om)=\frac{1}{2N}\sum_{\bk b\xi}
\frac{\bigl[(\om+\te^b_{\bk\xi})(\om+\te^b_{\bk-\bq\xi})-\Delta^b_{\bk\xi}\Delta^b_{\bk-\bq\xi}\bigr]
\bigl[(\om)^2-\tE^{b2}_{\bk \bxi}\bigr]\bigl[(\om)^2-\tE^{b2}_{\bk-\bq\bxi}\bigr]}
{[(\om)^2-\Om^{b2}_{\bk +}][(\om)^2-\Om^{b2}_{\bk-}][(\om)^2-\Om^{b2}_{\bk-\bq +}][(\om)^2-\Om^{b2}_{\bk-\bq-}]},
\label{eq:QPI3Dex}
\ee
\hfill\noindent\rule{0.5\linewidth}{0.4pt}
\twocolumngrid
~\newline
where $\xi=\pm$ and $\bxi=-\xi$. The denominator  in  Eq.~(\ref{eq:QPI3Dex}) is equal to the product $D_b(\bk,\om)D_b(\bk-\bq,\om)$.
The above expression for $\Lambda_0(\bq,\om)$ reduces to  the 2D expression in Eq.~(\ref{eq:QPI2D}) for $\varep^{b}_{c\bk}=0$.
In contrast to Eq.~(\ref{eq:QPI2D}) the momentum integral also includes the $k_z$- direction in Eq.~(\ref{eq:QPI3Dex}) .
In the above expressions   for $\Lambda_0(\bq,\om)$ we have neglected terms $\sim |\varep^b_{c\bk}|^2, |\varep^b_{c\bk}|^4$ in the numerators since they influence only the amplitude.\\

Now we discuss the numerical results for the expected QPI spectrum calculated with Eq.~(\ref{eq:QPI3Dex}). It turns out that the influence of the c-axis dispersion in the bands is of little importance due to the smallness of $t_c^b$ in the present case of \SP. Although small differences in the 3D QPI contribution of each individual $k_z$ slice are present, the integration along $k_z$ smoothes the differences to the simple 2D case described by Eq.~(\ref{eq:QPI2D}). 

%
\begin{figure}[t]
\vspace{0.cm}
\includegraphics[width=0.98\linewidth]{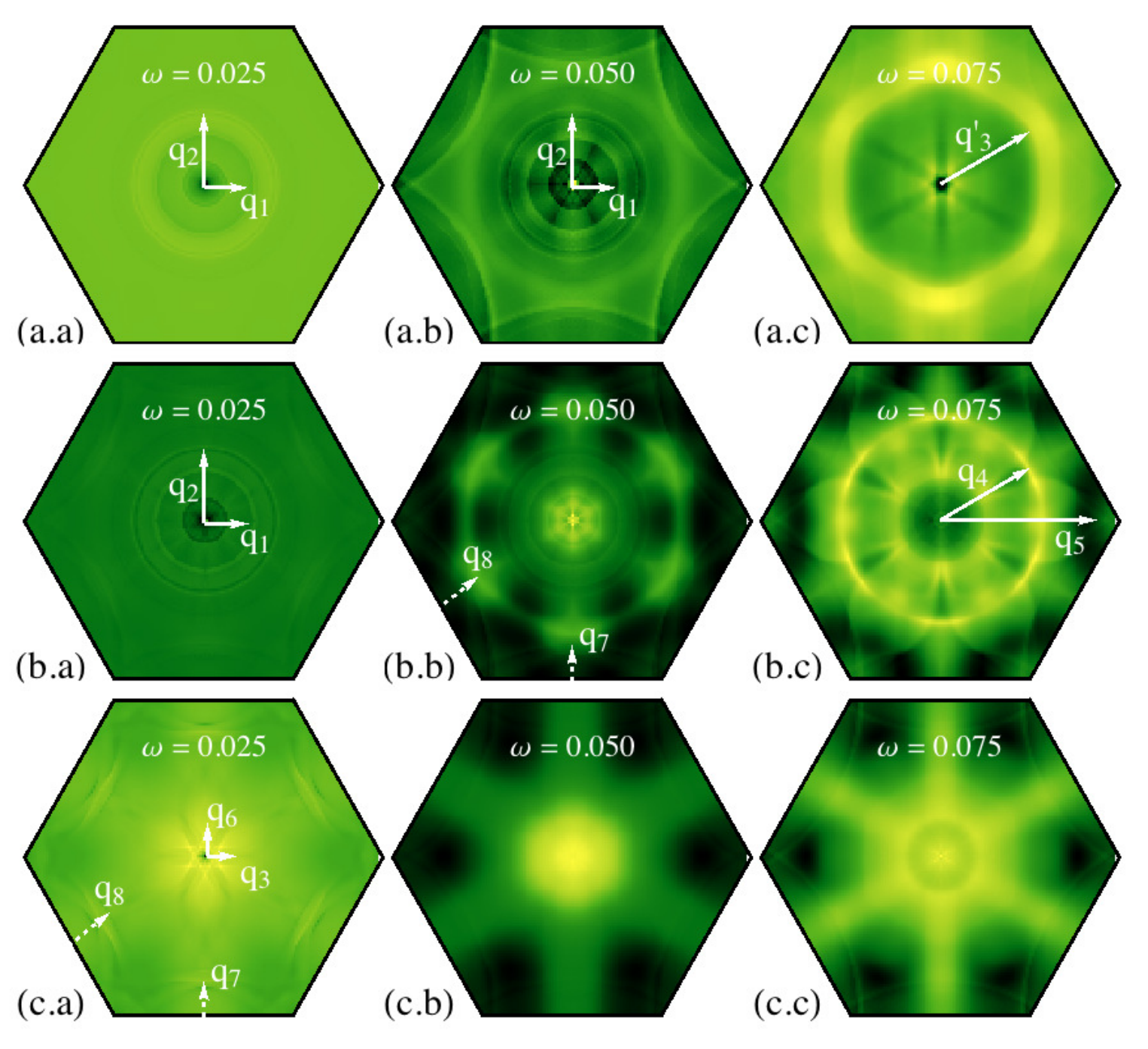}
\caption{ (Color online)
QPI spectrum (average) for the three gap candidates $A_{1g}$ (top row (a,a-c));  $A_{2u}$ (middle row (b,a-c) and 
$E_{g}$ (bottom row (c,a-c)) for three frequencies. The characteristic wave vectors $\bq_i$ of the equal energy SC quasiparticle
surfaces in Fig.~\ref{fig:Fig2}d-f can be identified in the structure of some QPI spectra.}
\label{fig:Fig4}
\end{figure}
%

We first consider the normal state whose two  cuts of spectral functions (2D)  are shown in Fig.~(\ref{fig:Fig1}). Particularly, in Fig.~(\ref{fig:Fig1}.b) the Fermi surface is plotted  for $k_z=\pi/c$ with the typical characteristic wave vectors denoted  by 
$\bq_i' $~$(i=1-6)$. The normal state DOS at the Fermi level is  $74\%$ of band-3 character \cite{goryo:12}. The QPI  should therefore be dominated by this band. Indeed this is found when considering the individual $b=1-3$ contributions in Eq.~(\ref{eq:QPI2D}). The two main features in  Fig.~\ref{fig:Fig3}(a) are a large central ring and touching arcs around  the zone boundary (K) points.
 The ring is due to $\bq'_3$ and $\bq_6'$ scattering (Fig.~\ref{fig:Fig1}b) inside and between band-3 sheets (the dashed arrows are folded back into the first BZ). The arcs are due to $\bq'_4$ type scattering between different band-3 sheets.
When the voltage increases the ring shrinks due to the hole type bands. In addition linear features 
perpendicular to the hexagonal sides appear. They are due to a continuum of  $\bq'_3$-$\bq'_6$ scattering with the result of the averaging over the different $k_z$ cuts.

The superconducting candidate states have very different nodal structure (Fig.~\ref{fig:Fig2}a-c) and therefore also different quasiparticle equal energy surfaces and associated characteristic scattering wave vectors $\bq_i$~$(i=1-8)$ (Fig.~\ref{fig:Fig2}.d-f).
This leads to three distinct QPI spectra for the gap candidates shown in  Fig.~\ref{fig:Fig4}. They also exhibit a considerably different behavior as function of bias voltage or frequency. A few characteristic wave vectors $\bq_i$ associated with the equal energy surfaces in (Fig.~\ref{fig:Fig2}.d-f) can clearly be seen in the QPI spectrum of  Fig.~\ref{fig:Fig4} for low frequencies. In particular the faint rings with $\bq_1,\bq_2$ due to the small $b=2$ band are now visible in Fig.~\ref{fig:Fig4}(a.a) and (b.a) because the contribution of the b=3 band is mostly gapped out for $A_{1g}$ and $A_{2u}$. For $E_{g}$ in   Fig.~\ref{fig:Fig4}(c.a) however the different node structure leads to particular  scattering wave vectors ($\bq_{4-8}$), on $b=1$ sheets. In principle, $\bq_{7,8}$  resemble the normal state $\bq'_{1,2}$. For larger $\omega$ they also appear for $A_{2u}$ in  Fig.~\ref{fig:Fig4}(b.b-c). At still larger $\omega=0.075$ some features of the normal state QPI at $\bq'_3$ reappear in Fig.~\ref{fig:Fig4}(a-c.c). Also the scattering between different $b=3$ sheets perpendicular  to hexagonal BZ directions  appear in the E$_g$ QPI of Fig.~\ref{fig:Fig4}(c.c).

To summarize we have presented the QPI theory in Born approximation for hexagonal pnictide superconductor \SP. Its main hole band can be clearly identified in the normal state QPI. In the superconducting state the three candidate gap functions proposed in Ref.~\cite{goryo:12} show different types of  equal energy quasiparticle sheets leading to three distinct QPI pattern and bias voltage dependences. Therefore a detailed experimental investigation of QPI in \SP~ should be able to discriminate between the theoretically proposed gap symmetries. This is particularly desirable because recent NMR and NQR experiments \cite{matano:14} suggest a fully gapped spin singlet state.


\bibliography{References}
\bibliographystyle{arXiv}
\end{document}